\documentclass[aps,twocolumn,showpacs]{revtex4}
\usepackage{graphicx}
\usepackage{dcolumn}
\usepackage{amsmath}
\usepackage{enumerate}

\begin{document}

\title{Spin-tensor decomposition of nuclear transition matrix elements for
neutrinoless double-$\beta $ decay of $^{76}$Ge and $^{82}$Se nuclei within
PHFB approach}

\author{P. K. Rath$^{1}$}
\author {R. Chandra$^{2}$}
\thanks {Corresponding author: ramesh.luphy@gmail.com}
\author {A. Kumar$^{1}$}
\author {P. K. Raina$^{3}$}
\author {B. M. Dixit$^{4}$}
\affiliation {
$^{1}$Department of Physics, University of Lucknow, Lucknow-226007, India}
\affiliation {
$^{2}$Department of Applied Physics, Babasaheb Bhimrao Ambedkar University,
Lucknow-226025, India}
\affiliation {
$^{3}$Department of Physics, Indian Institute of Technology, Ropar,
Rupnagar - 140001, India}
\affiliation{
$^{4}$Faculty of Physical Sciences, SRM University, Barabanki-225013, India}

\date{\today}

\begin{abstract}
Employing the PHFB model, nuclear transition matrix elements $M^{\left(
K\right) }$ for the neutrinoless double-$\beta^{-} $ decay of $\ ^{76}$Ge 
and $^{82}$Se isotopes are calculated within mechanisms involving light 
as well as heavy Majorana neutrinos, and classical Majorons by considering 
the spin-tensor decomposition of realistic KUO and empirical JUN45 effective 
two-body interaction. It is noticed that the effects due to the SRC on 
NTMEs $M^{\left( 0\nu \right) }$ and $M^{\left( 0N\right) }$ due to the 
exchange of light and heavy Majorana neutrinos, respectively, is maximally 
incorporated by the central  part of the effective two-body interaction, 
which varies by a small amount with the inclusion of spin-orbit and tensor 
components. The maximum uncertainty in the average NTMEs $\overline{M}^{(0\nu)}$ 
and $\overline{M}^{(0N)}$ turns out to be about 10\% and 37\%, respectively.
\end{abstract}

\pacs{21.60.Jz,23.20.-g,23.40.Hc}

\maketitle

\section{INTRODUCTION}
                                                                          
In any gauge theoretical model, the possible occurrence of neutrinoless double 
beta ($0\nu \beta \beta $) decay is intimately associated with the violation of 
lepton number $L$ conservation. Out of several possible mechanisms involving 
left-right symmetry, $R_{p}$-violating suspersymmetry, Majorons, sterile neutrinos, 
leptoquarks, compositeness and extra-dimensional scenarios \cite{verg16,dell16}, 
the Majorana neutrino mass mechanism is considered as the standard one to ascertain 
the Dirac or Majorana nature of neutrinos. The half-lives $T_{1/2}^{0\nu }$ of 
the $0\nu\beta \beta $ decay is a product of phase space factors, nuclear 
transition matrix elements (NTMEs) and corresponding gauge theoretical parameters. 
As the phase space factors have been calculated to good accuracy in the recent
past \cite{koti12,stoi13,simk15}, the accuracy of the extracted limits on
the parameters of a particular gauge theoretical model depends on the
reliability of model dependent NTMEs. Specifically, the effective mass of
the light and heavy Majorana neutrinos are extracted in the standard mass
mechanism. Over the past years, the theoretical studies devoted to the
calculation of NTMEs have been excellently reviewed in Ref. \cite{enge16}
and references there in.

In the evaluation of NTMEs, different theoretical approaches, namely
interacting shell-model (ISM) calculations based on direct diagonalization 
\cite{caur08,mene09,horo14,senk16}, QRPA \cite{voge86,civi87} and its
extensions \cite{suho98,faes98}, deformed QRPA, \cite{faes12,must13}, QRPA
with isospin restoration \cite{simk13}, projected-Hartree-Fock-Bogoliubov
(PHFB) \cite{rath10,rath13,rath12,rath16}, interacting boson model (IBM) 
\cite{bare13,yosi13} with isospin restoration \cite{bare15}, the generator 
coordinate method (GCM) \cite{rodr10}, and beyond mean field covariant density
functional theory (BMFCDFT)\ \cite{yao014} have been employed. In spite of
the fact that, several alternatives are available for the choice of model
space, effective two-body residual interactions, model dependent form
factors to include the finite size of nucleons (FNS), short range
correlations (SRC) with Miller-Spencer parametrization \cite{mill76},
unitary operator method (UCOM) \cite{kort07} parametrization based on
coupled cluster method (CCM) \cite{simk09}, and the value of axial vector
current coupling constant $g_{A}$ \cite{mend11,suho13,enge14,bare15}, the
calculated NTMEs $M^{(0\nu )}$ interestingly differ by factor of 2--3.

In addition to these exciting developments in the theoretical front, the
remarkable experimental studies of the $\beta \beta $ decay \cite{saak13} 
have resulted in measuring half lives $T_{1/2}^{0\nu }$ of 
$0\nu \beta^{-} \beta ^{-}$ decay of $^{76} $Ge, $^{100}$Mo, $^{130}$Te 
and $^{136}$Xe isotopes to be $>$3.0 $\times $10$^{25}$ yr by the combined
data of the Heidelberg-Moscow experiment \cite{klap01}, 
international germanium experiment (IGEX) \cite{aals02} and GREDA-I \cite{agos13}, 
$>$ 1.1 $\times $10$^{24}$ yr by NEMO-3 \cite{bara11}, $>$ 4.0 $\times $10$^{24}$ 
yr by CUORE \cite{alfo15} and $>$ 1.1 $\times $10$^{26}$ yr by KamLAND-Zen 
\cite{gand16} ($>$ 1.6 $\times $10$^{25}$ by EXO \cite{auge12}), respectively. 
Our present concern is to calculate
NTMEs for the $0\nu \beta^{-}\beta^{-}$ decay of $^{76}$Ge, and $^{82}$Se
isotopes, which in turn requires the reliable wave functions of $^{76}$Ge, $%
^{76,82}$Se and $^{82}$Kr nuclei. As the wave functions are model dependent,
the employed model should be versatile enough to reproduce as many observed
properties of nuclei as possible.

An important observed characteristic feature of nuclei in the Ge region is the
shape transitions at $N=40$. The onset of deformation at $N=40$ necessitates
to adopt a calculational framework treating the interplay of pairing and
deformation degrees of freedom simultaneously, and on equal footing \cite%
{good79}. Calculations have already been performed by using ISM in a
valance space spanned by the 1p$_{1/2}$, 1p$_{3/2}$, 0f$_{5/2}$ and 0g$%
_{9/2} $ orbits treating the doubly even $^{56}$Ni as an inert core. The
present calculation is performed employing the PHFB approach in the above
mentioned valance space with a realistic and an empirical two body effective
interaction, namely KUO \cite{kuo01} and JUN45 \cite{honm09}, respectively.
The purpose of such a calculation is to demonstrate the difference between
the two approaches and infer about the role of neglected configurations and
quasiparticle interactions.

\section{THEORETICAL\ FORMALISM}

The detailed theoretical formalism required for the study of the $0\nu 
\beta^{-} \beta^{-} $ decay within the Majorana neutrino mass mechanism
has been given in Refs. \cite{simk99,verg02}. Within
the PHFB model, the calculation of NTMEs due to the exchange of light \cite%
{rath13} and heavy Majorana \cite{rath12} neutrinos has already been
reported. In the following, we present a brief out line of the required
formalism for the clarity in notations used in the present paper.

Within the Majorana neutrino mass mechanism, the half-life $T_{1/2}^{\left(
0\nu \right) }$ for the 0$^{+}\rightarrow $0$^{+}$\ transition of 
$0\nu \beta^{-} \beta^{-} $ decay is given by

\begin{equation}
\left[ T_{1/2}^{\left( 0\nu \right) }(0^{+}\rightarrow 0^{+})\right]
^{-1}=G_{01}\left\vert \frac{\left\langle m_{\nu }\right\rangle }{m_{e}}%
M^{\left( 0\nu \right) }+\frac{m_{p}}{\left\langle M_{N}\right\rangle }%
M^{\left( 0N\right) }\right\vert ^{2}
\end{equation}%
where 
\begin{eqnarray}
\left\langle m_{\nu }\right\rangle &=&\sum\nolimits_{i}^{\prime
}U_{ei}^{2}m_{i},\qquad \qquad m_{i}<10\text{ eV}, \\
\left\langle M_{N}\right\rangle ^{-1} &=&\sum\nolimits_{i}^{\prime \prime
}U_{ei}^{2}m_{i}^{-1},\qquad \qquad m_{i}>10\text{ GeV}, \\
M^{\left( 0K\right) } &=&-\frac{M_{F}^{\left( 0K\right) }}{g_{A}^{2}}%
+M_{GT}^{\left( 0K\right) }+M_{T}^{\left( 0K\right) }
\end{eqnarray}%
and the $K=$ $0\nu $ $\left( 0N\right) $ denotes mass mechanism due to the
exchange of light (heavy) Majorana neutrinos. The phase space factors

\begin{equation}
G_{01}=\left[ \frac{2\left( G_{F}\text{ }g_{A}\right) ^{4}m_{e}^{9}}{64\pi
^{5}\text{ }\left( m_{e}R\right) ^{2}\ln \left( 2\right) }\right]
\int_{1}^{T+1}f_{11}^{(0)}p_{1}\text{ }p_{2}\text{ }\varepsilon
_{1}\varepsilon _{2}\text{ }d\varepsilon _{1}
\end{equation}%
have been recently calculated with good accuracy incorporating the screening
correction \cite{koti12,stoi13,simk15} and the calculation of the NTMEs $%
M^{\left( K\right) }$ of the $0\nu \beta^{-} \beta^{-} $ decay within the 
PHFB model, has already been discussed in Ref. \cite%
{rath10,rath13,rath12}.

Employing the HFB wave functions, one obtains the following expression for
the NTME $M_{\alpha }^{\left( K\right) }$ of the $0\nu \beta^{-} \beta^{-} $ 
decay corresponding to an operator $O_{\alpha }^{(K)}$ 
\cite{rath10}.
\begin{eqnarray}
M_{\alpha }^{\left( K\right) }  &=&\left\langle 0_{f}^{+}\left\Vert O_{\alpha
}^{(K)}\right\Vert 0_{i}^{+}\right\rangle  \nonumber \\
&=&\left[ n^{Ji=0}n^{J_{f}=0}\right]
^{-1/2}  \nonumber \\
&&\times \int\limits_{0}^{\pi }n_{(Z,N),(Z+2,N-2)}(\theta
)\sum\limits_{\alpha \beta \gamma \delta }\left ( \alpha \beta \left|
O_{\alpha }^{(K)}\right| \gamma \delta \right)   \nonumber \\
&&\times \sum\limits_{\varepsilon \eta }\frac{\left( f_{Z+2,N-2}^{(\pi
)*}\right) _{\varepsilon \beta }}{\left[ \left( 1+F_{Z,N}^{(\pi )}(\theta
)f_{Z+2,N-2}^{(\pi )*}\right) \right] _{\varepsilon \alpha }}  \nonumber
\\
&&\times \frac{\left( F_{Z,N}^{(\nu )*}\right) _{\eta \delta }}{\left[
\left( 1+F_{Z,N}^{(\nu )}(\theta )f_{Z+2,N-2}^{(\nu )*}\right) \right]
_{\gamma \eta }}sin\theta d\theta ,  \label{ntm}
\end{eqnarray}
where
\begin{eqnarray}
n^{J} &=&\int\limits_{0}^{\pi }\left[ det\left( 1+F^{(\pi )}f^{(\pi
)^{\dagger }}\right) \right] ^{1/2}  \nonumber \\
&&\times \left[ det\left( 1+F^{(\nu )}f^{(\nu )^{\dagger }}\right) \right]
^{1/2}d_{00}^{J}(\theta )sin(\theta )d\theta ,
\end{eqnarray}

The required amplitudes $(u_{im},v_{im})$ and expansion coefficients $%
C_{ij,m}$ of axially symmetric HFB intrinsic state ${|\Phi _{0}\rangle }$
with $K=0$ to evaluate the expressions $n^{J}$, $n_{(Z,N),(Z+2,N-2)}(\theta
) $, $f_{Z,N}$\ \ and $F_{Z,N}(\theta )\ $\cite{rath10}  are obtained by
minimizing the expectation value of the effective Hamiltonian in a basis
constructed by using a set of deformed states.

\begin{table*}[htbp]
\caption{Comparison of calculated and observed excited energies $E_{2^{+}}$ 
of yrast 2$^{+}$ states \cite{saka84}, deformation parameters $\beta _{2}$ 
\cite{rama01} and g-factors g(2$^{+}$) \cite{ragh89} of $^{76}$Ge, $^{76,82}$Se 
and $^{82}$Kr isotopes with (a) HFB1 and (b) HFB2.}
\label{tab1}
\begin{tabular}{ccccccccccccccccc}
\hline\hline
{\small Nuclei} &  &  &  & \multicolumn{3}{c}{{\small E}$_{2+}$} &  &  &
\multicolumn{3}{c}{${\small \beta }_{2}$} &  &  & \multicolumn{3}{c}{\small %
g-factor} \\ \cline{5-7}\cline{10-12}\cline{15-17}
&~~~~~~  &~~~~~  &~~~~~~  & {\small Theo.} &~~~~~~~  & {\small Expt.} &~~~~~  &~~~~~  & {\small Theo.} &~~~~~~~  &
{\small Expt.} &~~~~~  &~~~~~  & {\small Theo.} &~~~~~  & {\small Expt.} \\ \hline
$^{76}${\small Ge} &  & {\small (a)} &  & {\small 0.563} &  & {\small 0.563}
&  &  & {\small 0.2610} &  & ${\small 0.2623\pm 0.0039}$ &  &  & {\small %
0.353 (0.353)} &  & {\small 0.334}$\pm 0.038$ \\
&  & {\small (b)} &  & {\small 0.535} &  &  &  &  & {\small 0.2483} &  &  &
&  & {\small 0.299 (0.306)} &  &  \\
$^{76}${\small Se} &  & {\small (a)} &  & {\small 0.559} &  & {\small 0.559}
&  &  & {\small 0.2991} &  & ${\small 0.3090\pm 0.0037}$ &  &  & {\small %
0.394 (0.394)} &  & {\small 0.40}$\pm 0.11$ \\
&  & {\small (b)} &  & {\small 0.507} &  &  &  &  & {\small 0.3004} &  &  &
&  & {\small 0.305 (0.322)} &  &  \\
$^{82}${\small Se} &  & {\small (a)} &  & {\small 0.659} &  & {\small 0.654}
&  &  & {\small 0.1988} &  & ${\small 0.1934\pm 0.0037}$ &  &  & {\small %
0.580 (0.522)} &  & {\small 0.43}$\pm 0.12$ \\
&  & {\small (b)} &  & {\small 0.641} &  &  &  &  & {\small 0.2142} &  &  &
&  & {\small 0.367 (0.357)} &  &  \\
$^{82}${\small Kr} &  & {\small (a)} &  & {\small 0.767} &  & {\small 0.777}
&  &  & {\small 0.2048} &  & ${\small 0.2021\pm 0.0045}$ &  &  & {\small %
0.489 (0.467)} &  &  \\
&  & {\small (b)} &  & {\small 0.779} &  &  &  &  & {\small 0.2028} &  &  &
&  & {\small 0.448 (0.436)} &  &  \\ \hline\hline
\end{tabular}
\end{table*}

\begin{table*}[htbp]
\caption{Calculated (Theo.) and observed (Expt.) occupation numbers for
neutrons \cite{schi08} and protons \cite{kay09} in $^{76}$Ge, $^{76,82}$Se
and $^{82}$Kr isotopes with (a) HFB1 and (b) HFB2.}
\label{tab2}
\begin{tabular}{cccccccccccccccccc}
\hline\hline
{\small Orbits} &  &  &  & \multicolumn{3}{c}{$^{76}${\small Ge}} &  &  &
\multicolumn{3}{c}{$^{76}${\small Se}} &  &  & $^{82}${\small Se} &  &  & $%
^{82}${\small Kr} \\ \cline{5-7}\cline{10-12}
&~~~~~~  &~~~~~~  &~~~~~~  & {\small Theo.} &~~~~~~~  & {\small Expt.} &~~~~~~  &~~~~~~  & {\small Theo.} &~~~~~~~  &
{\small Expt.} &~~~~~~  &~~~~~~  & {\small Theo.} &~~~~~~  &~~~~~~  & {\small Theo.} \\ \hline
{\small Protons} &  &  &  &  &  &  &  &  &  &  &  &  &  &  &  &  &  \\
{\small 1}$p_{1/2}+${\small 1}$p_{3/2}$ &  & {\small (a)} &  & {\small 1.60}
&  & {\small 1.75}$\pm 0.15$ &  &  & {\small 2.54} &  & {\small 2.09}$\pm
0.15$ &  &  & {\small 1.98} &  &  & {\small 2.98} \\
&  & {\small (b)} &  & {\small 0.96} &  &  &  &  & {\small 1.88} &  &  &  &
& {\small 1.98} &  &  & {\small 2.37} \\
{\small 0}$f_{5/2}$ &  & {\small (a)} &  & {\small 2.30} &  & {\small 2.04}$%
\pm 0.25$ &  &  & {\small 3.20} &  & {\small 3.17}$\pm 0.25$ &  &  & {\small %
3.55} &  &  & {\small 3.43} \\
&  & {\small (b)} &  & {\small 2.77} &  &  &  &  & {\small 3.42} &  &  &  &
& {\small 3.75} &  &  & {\small 3.86} \\
{\small 0}$g_{9/2}$ &  & {\small (a)} &  & {\small 0.10} &  & {\small 0.23}$%
\pm 0.25$ &  &  & {\small 0.26} &  & {\small 0.86}$\pm 0.25$ &  &  & {\small %
0.47} &  &  & {\small 1.58} \\
&  & {\small (b)} &  & {\small 0.26} &  &  &  &  & {\small 0.70} &  &  &  &
& {\small 0.27} &  &  & {\small 1.76} \\ \hline
{\small Neutrons} &  &  &  &  &  &  &  &  &  &  &  &  &  &  &  &  &  \\
{\small 1}$p_{1/2}+${\small 1}$p_{3/2}$ &  & {\small (a)} &  & {\small 4.95}
&  & {\small 4.87}$\pm 0.20$ &  &  & {\small 4.09} &  & {\small 4.41}$\pm
0.20$ &  &  & {\small 5.76} &  &  & {\small 5.51} \\
&  & {\small (b)} &  & {\small 4.48} &  &  &  &  & {\small 3.75} &  &  &  &
& {\small 5.97} &  &  & {\small 5.41} \\
{\small 0}$f_{5/2}$ &  & {\small (a)} &  & {\small 4.24} &  & {\small 4.56}$%
\pm 0.40$ &  &  & {\small 3.73} &  & {\small 3.83}$\pm 0.20$ &  &  & {\small %
5.72} &  &  & {\small 5.05} \\
&  & {\small (b)} &  & {\small 4.73} &  &  &  &  & {\small 3.96} &  &  &  &
& {\small 5.97} &  &  & {\small 5.49} \\
{\small 0}$g_{9/2}$ &  & {\small (a)} &  & {\small 6.81} &  & {\small 6.48}$%
\pm 0.30$ &  &  & {\small 6.18} &  & {\small 5.80}$\pm 0.30$ &  &  & {\small %
8.52} &  &  & {\small 7.45} \\
&  & {\small (b)} &  & {\small 6.79} &  &  &  &  & {\small 6.29} &  &  &  &
& {\small 8.05} &  &  & {\small 7.10} \\ \hline\hline
\end{tabular}
\end{table*}

\section{RESULT AND DISCUSSIONS}

Two different set of wave functions are generated using two distinct
effective interactions, namely KUO \cite{kuo01} and JUN45 due to 
Honma \textit{et al.} \cite{honm09}. The former is a realistic interaction 
while the latter is an empirical one. The wave functions obtained by using 
KUO and JUN45 effective two-body interactions are referred to as HFB1 and 
HFB2, respectively. The single particle energies (SPE) used in HFB1 (HFB2)
calculation are $\varepsilon _{p_{3/2}}$ = 0.0 (-9.828) MeV, $\varepsilon
_{f_{5/2}}$ = 0.78 (-9.048) MeV, $\varepsilon _{p_{1/2}}$ = 1.08 (-8.7480)
MeV and $\varepsilon _{g_{9/2}}$ = 3.0 (-6.828) MeV. However, the SPE of $%
\varepsilon _{g_{9/2}}$ = 4.0 (-6.828) MeV for $^{76}$Ge isotope. Usually, a
mass dependent term of the type (58/A)$^{1/3}$ is introduced \cite{wild84}
in the effective two body interaction to compensate for the noticed over
attractiveness of the interaction for the nuclei with high neutron number
occurring towards the end of the shell \cite{trip84}. The above mentioned 
effective interactions, namely KUO and JUN45 are renormalized to reproduce 
the excitation energies $E_{2^{+}}$ of the yrast 2$^{+}$ states. 

To ascertain the reliability of the generated wave functions HFB1 and HFB2,
the calculated and experimentally observed excitation energies $E_{2^{+}}$ 
of the yrast 2$^{+}$ states \cite{saka84},
deformation parameters $\beta _{2}$ \cite{rama01} and g-factors g(2$^{+}$) 
\cite{ragh89} are presented in Table 1. The deformation parameters $\beta
_{2}$ are calculated with effective charges $e_{p}$=1+e$_{eff}$
and $e_{n}$=e$_{eff}$. The effective charge e$_{eff}$=0.5 for $^{82}$Kr
while for other nuclei it is 0.78. The g-factors g(2$^{+}$) are calculated
with two different prescriptions. In the first prescription, the g(2$^{+}$)
values are calculated with g$_{l}^{\pi }$ = 1.0, g$_{l}^{\nu }$=0.0, g$%
_{s,eff}^{\pi /\nu }$=0.6(g$_{s}^{\pi /\nu }$)$_{bare}$. In the second
prescription, effective operators calculated with a set of first and second
order diagrams are (g$_{l}$, g$_{s}$, g$_{p}$)$^{\pi }$ = (0.89, 3.18, 0.73)
and (g$_{l}$, g$_{s}$, g$_{p}$)$^{\nu }$ = (0.07, -1.52, -0.89) \cite{rath88}%
. In Table 2, the calculated occupation numbers are given along with the
experimentally observed data \cite{schi08,kay09}. It is noticed that the 
overall agreement between the calculated spectroscopic properties of $^{76}$Ge, $%
^{76,82}$Se and $^{82}$Kr isotopes and the experimentally observed data is
reasonably good.
Although the closure approximation is not valid for the 2$\nu \beta
^{-}\beta ^{-}$ decay, an estimate of  $M_{2\nu }$ with closure for $^{76}$%
Ge and $^{82}$Se provides 0.157 (0.132) and 0.155 (0.147) with HFB1 and
HFB2, respectively. This implies $g_{A,eff}=$0.667 (0.729) and 0.576 (0.592)
for $^{76}$Ge and $^{82}$Se isotopes with HFB1 and HFB2, respectively.

\begin{table*}[htbp]
\caption{ NTMEs for the 0$\nu $ $\beta^{-} \beta^{-} $ decay of $^{76}$Ge 
and $^{82}$Se due to the light and heavy Majorana neutrino exchange with 
three sets of wave functions having central (C), central plus spin-orbit (CS) 
and central plus spin-orbit plus tensor (CST) for both (a) HFB1 and (b) HFB2.}
\label{tab3}
\begin{tabular}{ccccccccccccccc}
\hline\hline
{\small Nucleus} &  & {\small Case} &  &  &  & {\small HFB1} &  &  &  &  &
& {\small HFB2} &  &  \\ \cline{5-9}\cline{11-15}
&~~~~~~~~~~~~~~~  &~~~~~~~~~~  &~~~~~~~~~~  & {\small C} &~~~~~~~~~~  & {\small CS} &~~~~~~~~~~  & {\small CST} &~~~~~~~~~~  
& {\small C} &~~~~~~~~~~& {\small CS} &~~~~~~~~~~  & {\small CST} \\ \hline
\multicolumn{2}{c}{\small Light neutrino} &  &  &  &  &  &  &  &  &  &  &  &
&  \\
$^{76}${\small Ge} &  & \multicolumn{1}{l}{\small FNS} &  & {\small 1.574} &
& {\small 3.982} &  & {\small 5.628} &  & {\small 1.321} &  & {\small 3.560}
&  & {\small 5.346} \\
&  & \multicolumn{1}{l}{\small SRC1} &  & {\small 1.277} &  & {\small 3.490}
&  & {\small 4.858} &  & {\small 1.060} &  & {\small 3.024} &  & {\small %
4.507} \\
&  & \multicolumn{1}{l}{\small SRC2} &  & {\small 1.560} &  & {\small 3.945}
&  & {\small 5.564} &  & {\small 1.311} &  & {\small 3.515} &  & {\small %
5.270} \\
&  & \multicolumn{1}{l}{\small SRC3} &  & {\small 1.646} &  & {\small 4.087}
&  & {\small 5.785} &  & {\small 1.386} &  & {\small 3.669} &  & {\small %
5.511} \\
$^{82}${\small Se} &  & \multicolumn{1}{l}{\small FNS} &  & {\small 3.575} &
& {\small 5.991} &  & {\small 6.415} &  & {\small 2.872} &  & {\small 2.176}
&  & {\small 5.846} \\
&  & \multicolumn{1}{l}{\small SRC1} &  & {\small 3.003} &  & {\small 5.175}
&  & {\small 5.494} &  & {\small 2.371} &  & {\small 1.882} &  & {\small %
5.049} \\
&  & \multicolumn{1}{l}{\small SRC2} &  & {\small 3.542} &  & {\small 5.938}
&  & {\small 6.344} &  & {\small 2.843} &  & {\small 2.158} &  & {\small %
5.786} \\
&  & \multicolumn{1}{l}{\small SRC3} &  & {\small 3.706} &  & {\small 6.172}
&  & {\small 6.609} &  & {\small 2.986} &  & {\small 2.242} &  & {\small %
6.015} \\ \hline
\multicolumn{2}{c}{\small Heavy neutrino} & \multicolumn{1}{l}{} &  &  &  &
&  &  &  &  &  &  &  &  \\
$^{76}${\small Ge} &  & \multicolumn{1}{l}{\small FNS} &  & {\small 121.44}
&  & {\small 194.58} &  & {\small 298.33} &  & {\small 109.50} &  & {\small %
204.92} &  & {\small 320.41} \\
&  & \multicolumn{1}{l}{\small SRC1} &  & {\small 45.40} &  & {\small 69.89}
&  & {\small 104.34} &  & {\small 42.35} &  & {\small 70.36} &  & {\small %
110.10} \\
&  & \multicolumn{1}{l}{\small SRC2} &  & {\small 75.62} &  & {\small 118.68}
&  & {\small 179.84} &  & {\small 69.26} &  & {\small 122.77} &  & {\small %
191.67} \\
&  & \multicolumn{1}{l}{\small SRC3} &  & {\small 100.85} &  & {\small 160.12%
} &  & {\small 244.35} &  & {\small 91.52} &  & {\small 167.51} &  & {\small %
261.63} \\
$^{82}${\small Se} &  & \multicolumn{1}{l}{\small FNS} &  & {\small 256.85}
&  & {\small 322.09} &  & {\small 355.63} &  & {\small 233.28} &  & {\small %
119.80} &  & {\small 307.09} \\
&  & \multicolumn{1}{l}{\small SRC1} &  & {\small 110.74} &  & {\small 114.61%
} &  & {\small 123.04} &  & {\small 105.32} &  & {\small 44.88} &  & {\small %
105.49} \\
&  & \multicolumn{1}{l}{\small SRC2} &  & {\small 170.22} &  & {\small 196.09%
} &  & {\small 213.70} &  & {\small 157.98} &  & {\small 74.50} &  & {\small %
183.98} \\
&  & \multicolumn{1}{l}{\small SRC3} &  & {\small 218.65} &  & {\small 265.01%
} &  & {\small 291.01} &  & {\small 200.31} &  & {\small 99.38} &  & {\small %
250.99} \\ \hline\hline
\end{tabular}
\end{table*}

In addition, the two body effective interaction is further decomposed into
central (C), spin-orbit (S) and tensor (T) components \cite{kirs73} and the
effect of these components on NTMEs $M^{\left( K\right) }$ involved in 
0$\nu \beta^{-} \beta^{-}$ decay is studied. In spin-tensor
decomposition, the most general two-body interaction is written as%
\begin{eqnarray}
V(1,2) &=&\sum_{k=0,1,2}\left[ X^{(k)}\times S^{(k)}\right] ^{(0)}  \nonumber
\\
&=&\sum_{k=0,1,2}V^{(k)}
\end{eqnarray}%
where the most general two-particle spin operators are written as 
$ S_{1}^{(0)} = 1$, $S_{2}^{(0)}=\left[ \mathbf{\sigma }_{1}\times \mathbf{%
\sigma }_{2}\right] ^{(0)}$, $ S_{3}^{(1)}=\left[ \mathbf{\sigma }_{1}+%
\mathbf{\sigma }_{2}\right] ^{(1)}$, $ S_{4}^{(1)} = \left[ \mathbf{\sigma }_{1}-
\mathbf{\sigma }_{2}\right]^{(1)}$, $ S_{5}^{(1)}=\left[ \mathbf{\sigma }_{1}\times 
\mathbf{\sigma }_{2}\right] ^{(1)}$, and $S_{6}^{(2)}=\left[ \mathbf{\sigma }_{1}\times 
\mathbf{\sigma }_{2}\right] ^{(2)}$.

The central and tensor part of the effective two-boy interaction are
represented by $V^{(0)}$\ and $V^{(2)}$, respectively. The $V^{(1)}$ term
contains the symmetric $S_{3}^{(1)}$ as well as antisymmetric $S_{4}^{(1)}$
and $S_{5}^{(1)}$ spin-orbit operators. Three sets of wave functions are
generated with central (C), central plus spin-orbit (CS) and central plus
spin orbit plus tensor (CST) parts of the effective two-body interaction,
which are subsequently employed to calculate the required NTMEs $M^{(K)}$.

Employing these reliable wave functions, various NTMEs, namely Fermi $%
M_{F}^{\left( K\right) }$, Gamow-Teller (GT) $M_{GT}^{\left( K\right) }$
consisting of $M_{GT-AA}^{\left( K\right) }$, $M_{GT-AP}^{\left( K\right) }$, 
$M_{GT-PP}^{\left( K\right) }$, $M_{GT-MM}^{\left( K\right) }$ and tensor $%
M_{T}^{\left( K\right) }$ consisting of $M_{T-AP}^{\left( 0\nu \right) }$, $%
M_{T-PP}^{\left( 0\nu \right) }$, $M_{T-MM}^{\left( 0\nu \right) }$ are
calculated with $g_{V}=1.0$, $g_{A}=1.2701$ \cite{beri12}, 
$\kappa =\mu _{p}-\mu _{n}=3.70$%
, $\Lambda _{V}=0.850$ GeV and $\Lambda _{A}=1.086$ GeV. Three sets of NTMEs
are calculated by considering a Jastrow type of SRC simulating the effects
of Argonne V18 and CD-Bonn potentials in the self-consistent coupled cluster
method (CCM) \cite{simk09}, given by

\begin{equation}
f(r)=1-ce^{-ar^{2}}(1-br^{2})
\end{equation}%
where $a=1.1$ $fm^{-2}$, $1.59$ $fm^{-2}$, $1.52$ $fm^{-2}$, $b=0.68$ $%
fm^{-2}$, $1.45$ $fm^{-2}$, $1.88$ $fm^{-2}$ and $c=1.0$, $0.92$, $0.46$ for
Miller--Spencer parametrization, Argonne NN, CD-Bonn Potentials, and
are denoted by SRC1, SRC2 and SRC3, respectively. In Table III, the NTMEs $%
M_{{}}^{\left( 0\nu \right) }$ and $M_{{}}^{\left( 0N\right) }$ due to the
exchange of light and heavy Majorana neutrinos, respectively, are displayed.

\begin{table*}[htbp]
\caption{Relative changes (in \%) of the NTMEs $M^{(0\nu )}$ ($M^{(0N)}$%
) with the inclusion of SRC (SRC1, SRC2, and SRC3), and average energy
denominator $A/2$.}
\label{tab4}
\begin{tabular}{cccccccccccccccccc}
\hline\hline
{\small Nuclei} &  & {\small Cases} &  &  &  &  & {\small HFB1} &  &  &  &
&  &  &  & {\small HFB2} &  &  \\ \cline{6-10}\cline{14-18}
&~~~~  &~~~~  &~~~  &~~~  & {\small C} &~~~~  & {\small CS} &~~~~  & {\small CST} &~~~  &~~~  &~~~  &~~~
{\small C} &~~~~  & {\small CS} &~~~~  & {\small CST} \\ \hline
$^{76}${\small Ge} &  & \multicolumn{1}{l}{\small SRC1} &  &  & {\small 18.9
(62.6)} &  & {\small 12.4 (64.1)} &  & {\small 13.7 (65.0)} &  &  &  &
{\small 19.8 (61.3)} &  & {\small 15.0 (65.7)} &  & {\small 15.7 (65.6)} \\
&  & \multicolumn{1}{l}{\small SRC2} &  &  & {\small 0.9 (37.7)} &  &
{\small 0.9 (39.0)} &  & {\small 1.1 (39.7)} &  &  &  & {\small 0.8 (36.8)}
&  & {\small 1.3 (40.1)} &  & {\small 1.4 (40.2)} \\
&  & \multicolumn{1}{l}{\small SRC3} &  &  & {\small 4.5 (17.0)} &  &
{\small 2.6 (17.7)} &  & {\small 2.8 (18.1)} &  &  &  & {\small 4.9 (16.4)}
&  & {\small 3.0 (18.3)} &  & {\small 3.1 (18.3)} \\
&  & \multicolumn{1}{l}{{\small SRC1(}$A/2${\small )}} &  &  & {\small 6.2}
&  & {\small 9.7} &  & {\small 9.4} &  &  &  & {\small 5.6} &  & {\small 8.7}
&  & {\small 8.7} \\
&  & \multicolumn{1}{l}{{\small SRC2(}$A/2${\small )}} &  &  & {\small 6.0}
&  & {\small 9.1} &  & {\small 8.8} &  &  &  & {\small 5.4} &  & {\small 8.2}
&  & {\small 8.2} \\
&  & \multicolumn{1}{l}{{\small SRC3(}$A/2${\small )}} &  &  & {\small 5.9}
&  & {\small 8.9} &  & {\small 8.6} &  &  &  & {\small 5.3} &  & {\small 8.0}
&  & {\small 8.0} \\
$^{82}${\small Se} &  & \multicolumn{1}{l}{\small SRC1} &  &  & {\small 16.0
(56.9)} &  & {\small 13.6 (64.4)} &  & {\small 14.4 (65.4)} &  &  &  &
{\small 17.4 (54.8)} &  & {\small 13.5 (62.5)} &  & {\small 13.6 (65.6)} \\
&  & \multicolumn{1}{l}{\small SRC2} &  &  & {\small 0.9 (33.7)} &  &
{\small 0.9 (39.1)} &  & {\small 1.1 (39.9)} &  &  &  & {\small 1.0 (32.3)}
&  & {\small 0.8 (37.8)} &  & {\small 1.0 (40.1)} \\
&  & \multicolumn{1}{l}{\small SRC3} &  &  & {\small 3.7 (14.9)} &  &
{\small 3.0 (17.7)} &  & {\small 3.0 (18.2)} &  &  &  & {\small 4.0 (14.1)}
&  & {\small 3.1 (17.0)} &  & {\small 2.9 (18.3)} \\
&  & \multicolumn{1}{l}{{\small SRC1(}$A/2${\small )}} &  &  & {\small 6.4}
&  & {\small 9.4} &  & {\small 9.5} &  &  &  & {\small 5.3} &  & {\small 9.0}
&  & {\small 9.8} \\
&  & \multicolumn{1}{l}{{\small SRC2(}$A/2${\small )}} &  &  & {\small 6.2}
&  & {\small 8.8} &  & {\small 8.9} &  &  &  & {\small 5.3} &  & {\small 8.5}
&  & {\small 9.2} \\
&  & \multicolumn{1}{l}{{\small SRC3(}$A/2${\small )}} &  &  & {\small 6.1}
&  & {\small 8.6} &  & {\small 8.7} &  &  &  & {\small 5.2} &  & {\small 8.3}
&  & {\small 9.0} \\ \hline\hline
\end{tabular}
\end{table*}

Relative changes (in \%) of the NTMEs $M^{(0\nu )}$ ($M^{(0N)}$%
) with the inclusion of SRC1, SRC2, and SRC3, and average energy
denominator $A/2$ are given in Table IV.
In the case of light neutrino exchange, the contributions of C and CS parts
of the two-body interaction to the total NTMEs $M^{\left( 0\nu \right) }$ of
$^{76}$Ge calculated within HFB1 and HFB2 are  about 24\%--29\% and
67\%--72\%, respectively.  However, the contribution of the C part in the
case of $^{82}$Se turn out to be  47\% and 56\% for HFB1 and HFB2,
respectively. In the case of CS part, the contributions to the total NTMEs $%
M^{\left( 0\nu \right) }$ for HFB1 and HFB2 are  about 37\% and 94\%,
respectively. The maximum relative change in NTMEs $M^{\left( 0\nu \right) }$, when the
energy denominator is taken as $\overline{A}/2$ instead of $\overline{A}$
is of the order of 10 \%, which confirms that the dependence of NTMEs on the
average excitation energy $\overline{A}$ is small and thus, the validity of the
closure approximation for the  $0\nu \beta^{-} \beta^{-} $ decay is supported.
In comparison to the case FNS, the NTMEs $M^{\left( 0\nu \right) }$ are approximately
reduced by 14\%--16\%, 1\%--1.4\% and 2.8\%--3.0\% with the addition of
SRC1, SRC2 and SRC3, respectively.

The contributions of C and CS parts of the two-body interaction to the total
NTMEs $M^{\left( 0N\right) }$ of $^{76}$Ge due to the heavy neutrino
exchange, are about 34\%--42\% and 64\%--67\%, within HFB1 and HFB2,
respectively. However, the maximum contribution of the C part in the case of
$^{82}$Se turn out to be 94\% and 56\% for HFB1 and HFB2, respectively. In
the case of CS part, the contributions to the total NTMEs $M^{\left( 0\nu
\right) }$ for HFB1 and HFB2 are about 94\% and 43\%, respectively. With the
addition of SRC1, SRC2 and SRC3, the NTMEs $M^{\left( 0N\right) }$ are
approximately reduced by 64\%--66\%, 40\% and 15\%--18\% , respectively in
comparison to the case FNS. It is worth mentioning that the effects due to SRC 
on NTMEs $M^{\left( 0\nu \right) }$ and $M^{\left( 0N\right) }$ is maximally 
incorporated by the C part of the effective two-body interaction, which varies
by a small amount with the inclusion of spin-orbit and tensor parts.

\begin{table*}[htbp]
\caption{Extracted limits on the effective mass of neutrino ${\small <}%
m_{\nu }{\small >}$, ${\small <}M_{N}{\small >}$ and predicted half-lives $%
T_{1/2}^{(0\nu )}${\small \ (yrs)} with two sets of wave functions HFB1 and
HFB2 and (a) SRC1, (b) SRC2 and (c) SRC3.}
\label{tab5}
\begin{tabular}{cccccccccccccccccccccc}
\hline\hline
{\small Nuclei} &  & $T_{1/2}^{(0\nu )}${\small \ (yr)} & {\small Ref.} &  &
{\small SRC} &  & \multicolumn{3}{c}{$<m_{\nu }>${\small \ (eV)}} &  &  &  &
\multicolumn{3}{c}{{\small \ }$%
\begin{array}{c}
T_{1/2}^{(0\nu )}{\small \ (yr)} \\ 
({\small <}m_{\nu }{\small >=0.01\ eV)}%
\end{array}%
$} &  &  &  & \multicolumn{3}{c}{${\small <}M_{N}{\small >}${\small \ (GeV)}}
\\ \cline{8-10}\cline{14-16}\cline{20-22}
&~~~~  &~~~~  &~~~~  &~~~~  &~~~~  &~~~~  & {\small HFB1} &~~~~  & {\small HFB2} &~  &~~  &~~  & {\small HFB1}
&~~~~  & {\small HFB2} &~  &~~  &~~  & {\small HFB1} &~~~~  & {\small HFB2} \\ \hline
$^{76}${\small Ge} &  & {\small 3.0}$\times 10^{25}$ & {\small \cite{agos13}}
&  & {\small (a)} &  & {\small 0.24} &  & {\small 0.26} &  &  &  & {\small \
1.80}$\times 10^{28}$ &  & {\small 2.09}$\times 10^{28}$ &  &  &  & {\small %
4.20}$\times 10^{7}$ &  & {\small 4.44}$\times 10^{7}$ \\
&  &  &  &  & {\small (b)} &  & {\small 0.21} &  & {\small 0.22} &  &  &  &
{\small 1.37}$\times 10^{28}$ &  & {\small 1.53}$\times 10^{28}$ &  &  &  &
{\small 7.25}$\times 10^{7}$ &  & {\small 7.72}$\times 10^{7}$ \\
&  &  &  &  & {\small (c)} &  & {\small 0.20} &  & {\small 0.22} &  &  &  &
{\small 1.27}$\times 10^{28}$ &  & {\small 1.40}$\times 10^{28}$ &  &  &  &
{\small 9.85}$\times 10^{7}$ &  & {\small 1.05}$\times 10^{8}$ \\
$^{82}${\small Se} &  & {\small 3.6 }$\times ${\small \ 10}$^{23}$ & {\small
\cite{bara11}} &  & {\small (a)} &  & {\small 0.95} &  & {\small 1.04} &  &
&  & {\small 3.27}$\times 10^{27}$ &  & {\small 3.87}$\times 10^{27}$ &  &
&  & {\small 1.13}$\times 10^{7}$ &  & {\small 9.66}$\times 10^{6}$ \\
&  &  &  &  & {\small (b)} &  & {\small 0.82} &  & {\small 0.90} &  &  &  &
{\small 2.45}$\times 10^{27}$ &  & {\small 2.95}$\times 10^{27}$ &  &  &  &
{\small 1.96}$\times 10^{7}$ &  & {\small 1.68}$\times 10^{7}$ \\
&  &  &  &  & {\small (c)} &  & {\small 0.79} &  & {\small 0.87} &  &  &  &
{\small 2.26}$\times 10^{27}$ &  & {\small 2.73}$\times 10^{27}$ &  &  &  &
{\small 2.66}$\times 10^{7}$ &  & {\small 2.30}$\times 10^{7}$ \\
\hline\hline
\end{tabular}
\end{table*}

\begin{table}[htbp]
\caption{Extracted limits on the Majoron-neutrino coupling constant ${\small %
<}g_{M}{\small >}$ from the observed limits on ${\small T}_{1/2}^{\left(
0\nu \phi \right) }$(yr) with two sets of wave functions HFB1 and HFB2 and
(a) SRC1, (b) SRC2 and (c) SRC3.}
\label{tab6}
\begin{tabular}{cccccccccccc}
\hline\hline
{\small Nuclei} &  & $T_{1/2}^{\left( 0\nu \phi \right) }${\small (yr)} &  &
{\small Ref.} &  & {\small SRC} &  & \multicolumn{4}{c}{${\small <g}_{M}%
{\small >}$} \\ \cline{9-12}
&  &  &  &  &  &  &  & {\small HFB1} &~~  &  & {\small HFB2} \\ \hline
$^{76}${\small Ge} &  & {\small 6.4}$\times 10^{22}$ &  & {\small \cite%
{klap01}} &  & {\small (a)} &  & {\small 7.59}$\times 10^{-5}$ &  &  &
{\small 8.18}$\times 10^{-5}$ \\
&  &  &  &  &  & {\small (b)} &  & {\small 6.62}$\times 10^{-5}$ &  &  &
{\small 6.99}$\times 10^{-5}$ \\
&  &  &  &  &  & {\small (c)} &  & {\small 6.37}$\times 10^{-5}$ &  &  &
{\small 6.69}$\times 10^{-5}$ \\
$^{82}${\small Se} &  & {\small 1.5}$\times 10^{22}$ &  & {\small \cite%
{bara11}} &  & {\small (a)} &  & {\small 4.85}$\times 10^{-5}$ &  &  &
{\small 5.28}$\times 10^{-5}$ \\
&  &  &  &  &  & {\small (b)} &  & {\small 4.20}$\times 10^{-5}$ &  &  &
{\small 4.60}$\times 10^{-5}$ \\
&  &  &  &  &  & {\small (c)} &  & {\small 4.03}$\times 10^{-5}$ &  &  &
{\small 4.43}$\times 10^{-5}$ \\ \hline\hline
\end{tabular}
\end{table}

\begin{table}[htbp]
\caption{Extracted parameters from the observed limits on $T_{1/2}^{\left(
0\nu \right) }${\small \ and }$T_{1/2}^{\left( 0\nu \phi \right) }${\small \
}using average NTMEs $\overline{M}^{(0\nu )}$ and $\overline{M}^{(0N)}$.}
\label{tab7}
\begin{tabular}{lccccc}
\hline\hline
{\small Parameters} &~~~~~~~  & $^{76}${\small Ge} &~~~~  &~~~~~  & $^{82}${\small Se} \\
\hline
$\overline{M}^{(0\nu )}$ &  & {\small 5.249}$\pm 0.481$ &  &  & {\small 5.883%
}$\pm 0.568$ \\
${\small <}m_{\nu }{\small >}${\small \ (eV)} &  & {\small 0.227} &  &  &
{\small 0.890} \\
$T_{1/2}^{(0\nu )}${\small \ (yr)} &  & {\small 1.54}$\times 10^{28}$ &  &
& {\small 2.85}$\times 10^{27}$ \\
$\overline{M}^{(0N)}$ &  & {\small 181.99}$\pm 65.61$ &  &  & {\small 194.70}%
$\pm 72.13$ \\
${\small <}M_{N}{\small >}${\small \ (GeV)} &  & {\small 7.33}$\times 10^{7}$
&  &  & {\small 1.78}$\times 10^{7}$ \\
${\small <}g_{M}{\small >}$ &  & {\small 7.02}$\times 10^{-5}$ &  &  &
{\small 4.53}$\times 10^{-5}$ \\ \hline\hline
\end{tabular}
\end{table}

Limits on the effective neutrino mass $\left\langle m_{\nu }\right\rangle $
and $\left\langle M_{N}\right\rangle $ are extracted from the available
limits on experimental half-lives $T_{1/2}^{0\nu }$ using NTMEs $M^{(0\nu )}$
and $M^{(0N)}$calculated within the PHFB model (Table V). In the case of $^{76}$Ge
isotope using the HFB1 (HFB2) wave functions, one obtains the best limit on
the effective neutrino mass $\left\langle m_{\nu }\right\rangle <0.24$ eV,
0.21 eV, 0.20 eV (0.26 eV, 0.22 eV, 0.22 eV) and $\left\langle
M_{N}\right\rangle >4.20\times 10^{7}-9.85\times 10^{7}$ GeV (4.44$\times
10^{7}-10.5\times 10^{7}$ GeV) due to SRC1, SRC2 and SRC3, respectively. In
the classical Majoron model, the inverse half-life $T_{1/2}^{\left( 0\nu
\phi \right) }$ for the 0$^{+}\rightarrow $0$^{+}$ transition of Majoron
emitting $0\nu \beta ^{-}\beta ^{-}\phi$ decay is given
by \cite{doi85} 
\begin{equation}
\lbrack T_{1/2}^{\left( 0\nu \phi \right) }\left( 0^{+}\rightarrow
0^{+}\right) ]^{-1}=\left\vert \left\langle g_{M}\right\rangle \right\vert
^{2}.G_{\beta \beta \phi }.\left\vert M^{\left( 0\nu \phi \right)
}\right\vert ^{2}
\end{equation}%
where $\left\langle g_{M}\right\rangle $ is the effective Majoron-neutrino
coupling constant, and the NTME $M^{\left( 0\nu \phi \right) }$ is same as
the $M^{\left( 0\nu \right) }$ for the exchange of light Majorana neutrinos.
The phase space factors $G_{\beta \beta \phi }$ for the 0$^{+}\rightarrow $0$%
^{+}$ transition of $0\nu \beta ^{-}\beta ^{-}\phi$ decay
mode have been given by Kotila and Iachello \cite{koti15}. The extracted
limits on the effective Majoron-neutrino coupling parameter $\left\langle
g_{M}\right\rangle $ form the largest limits on the half-lives $%
T_{1/2}^{\left( 0\nu \phi \right) }$ are given in Table VI. The most
stringent extracted limit on $\left\langle g_{M}\right\rangle =\left(
6.37-8.18\right) \times 10^{-5}$.

In spite of the fact that there are only a set of six NTMEs $M^{\left( 0\nu
\right) }$ and $M^{\left( 0N\right) }$ for a statistical analysis to
estimate uncertainties therein, the calculated average NTMEs are given in
Table VII. The maximum uncertainty in the average NTMEs $\overline{M}^{(0\nu
)}$ and $\overline{M}^{(0N)}$turns out to be about 10\% and 37\%,
respectively.  Using the estimated average NTMEs $\overline{M}^{(0\nu )}$
and $\overline{M}^{(0N)}$ calculated in the PHFB model, the most stringent
extracted limits on the effective neutrino mass $\left\langle m_{\nu
}\right\rangle $ and $\left\langle M_{N}\right\rangle $ from the available
limit on experimental half-live $T_{1/2}^{0\nu }$ of $^{76}$Ge are 0.23 eV
and 7.33$\times 10^{7}$ GeV, respectively. Further, the extracted limit on
the effective Majoron-neutrino coupling parameter $\left\langle
g_{M}\right\rangle $ is $7.02\times 10^{-5}$.

\section{CONCLUSIONS}

Within the PHFB approach, the required NTMEs $M^{\left( 0K\right) }$ for the
study the $0\nu \beta^{-} \beta^{-} $ decay of $^{76}$Ge, and $^{82}$Se isotopes 
in the Majorana neutrino mass mechanism are
calculated using a two sets of HFB intrinsic wave functions, generated with
KUO and JUN45 effective two-body interactions. The reliability of the wave
functions has been tested by calculating the yrast spectra, deformation
parameter $\beta _{2}$ and and $g$-factors $g(2^{+})$ as well as $M_{2\nu }$
of nuclei participating in the $2\nu \beta^{-} \beta^{-} $
\ decay and comparing them with the available experimental data. An overall
agreement between the calculated and observed spectroscopic properties as
well as $M_{2\nu }$ suggests that the PHFB wave functions generated by
reproducing the $E_{2^{+}}$ are quite reliable. Further, the contributions
of the central, spin-orbit and tensor components of the effective two-body
interaction to the total $M^{\left( 0K\right) }$ have been obtained by
performing a spin-tensor decomposition of KUO and JUN45 two-body matrix
elements.

The NTMEs $M^{\left( 0\nu \right) }$ have a weak dependence on the average
excitation energy $\overline{A}$ of intermediate nucleus and as expected,
the closure approximation is quite valid.
In comparison to the case FNS, the NTMEs $M^{\left( 0\nu \right) }$
 ($M^{\left( 0N\right) }$) are approximately reduced by 15\% (65\%), 1\% (40\%)
and 3\% (16\%) with the addition of SRC1, SRC2 and SRC3, respectively.
Specifically, the strong dependence of $M^{\left( 0N\right) }$ in the case 
of heavy neutrino exchange on the SRC is a major source of uncertainty in 
the calculation of NTMEs. 
It has been noticed that the C part of the effective two-body interaction 
picks up maximally the effects due to SRC on NTMEs $M^{\left( 0\nu \right) }$ 
and $M^{\left( 0N\right) }$, which varies by a small amount with the inclusion 
of spin-orbit and tensor parts.

Limits on the effective light neutrino mass $%
\left\langle m_{\nu }\right\rangle $, effective heavy neutrino mass $%
\left\langle M_{N}\right\rangle $ and neutrino-Majoron coupling constant 
$\left\langle g_{M}\right\rangle $ of the classical Majoron model have been 
extracted from the available limits on experimental half-lives $T_{1/2}^{(0\nu )}$ 
and ${\small T}_{1/2}^{\left( 0\nu \phi \right) }$, respectively.
The most stringent extracted limits on $\left\langle m_{\nu}\right\rangle $,
 $\left\langle M_{N}\right\rangle $ and $\left\langle g_{M}\right\rangle $ 
from the available experimental limit on $T_{1/2}^{\left(0\nu\right) }$ of $^{76}$Ge are 0.23 eV
and 7.33$\times 10^{7}$ GeV, and $7.02\times 10^{-5}$, respectively.

\begin{acknowledgements}
This work is partially supported by DST-SERB, India vide grant 
no. SB/S2/HEP-007/2013, and Council of Scientific and Industrial 
Research (CSIR), India vide sanction No. 03(1216)/12/EMR-II.
\end{acknowledgements}

\end{document}